\documentstyle[11pt,cs10]{article}

\newcommand{\beq}{\begin{equation}}              
\newcommand{\eeq}{\end{equation}}             
\newcommand{\beqa}{\begin{eqnarray}}              
\newcommand{\eeqa}{\end{eqnarray}}             
\def\tR{{\tilde{R_s}}}
\def\etal{et al. }
\catcode`\@=11 
\def\lsim{\mathrel{\mathpalette\@versim<}}
\def\gsim{\mathrel{\mathpalette\@versim>}}
\def\@versim#1#2{\vcenter{\offinterlineskip
        \ialign{$\m@th#1\hfil##\hfil$\crcr#2\crcr\sim\crcr } }}
\catcode`\@=12 
\def\plottwo#1#2{\centering \leavevmode
a)\epsfxsize=.45\columnwidth \epsfbox{#1} \hfil
b)\epsfxsize=.45\columnwidth \epsfbox{#2}}

\begin{document}

\begin{flushright}
10$th$ Cambridge Workshop, Cool Stars, Stellar Systems and the Sun,\\
eds. R.Donahue \& J.Bookbinder, 1998.
\end{flushright}

\title{Surface Imaging by Microlensing}
 \author{Dimitar D. Sasselov}

\affil{Harvard-Smithsonian Center for Astrophysics, Cambridge, MA 02138,
USA}

\begin{abstract}
         Gravitational microlensing is a new technique for studying 
the surfaces of distant stars. A point mass lens, usually a low-mass
star from the disk, will typically resolve the surface of a red giant
in the Galactic bulge, as well as amplify its brightness by a factor
of 10 or more. Such events are now detected in real time and can be
followed up with precision photometry and spectroscopy.
The motivation for stellar imaging via
microlensing lies in its unique ability to provide center-to-limb
variation of spectral diagnostic lines. Such variation maps into a
variation with depth in the stellar atmosphere $-$  a unique and
very valuable constraint for modeling the structure of red giants'
atmospheres. We illustrate the technique on a recent successful
observational campaign and discuss the implications for current
stellar models.
\end{abstract}

\keywords{stellar imaging, microlensing, red giants, model atmospheres}

\section{Introduction}
Gravitational microlensing is a new technique for studying the surfaces
of distant stars (Sasselov 1996, 1997a), which is observationally viable
for objects in the Galactic bulge (Lennon \etal 1996, Alcock \etal 1997). 
Microlensing events have typical total durations of weeks to months.
While most lenses are point masses, many ($\sim$ 20\%) of the sources
monitored in the Galactic bulge are red giants and supergiants $-$
their angular radii are {\em comparable} to the Einstein
radius of a sub-solar mass lens (in the range of 50$\mu$as).
In addition to being resolved by most lenses, the projected disks of
bulge giants are large enough to make the probability for a lens transit
very high $-$ in fact, at least one such event (MACHO Alert 95-30) was
very well observed recently (Alcock \etal 1997), and two near misses occured
this summer (MACHO Alert 97-BLG-10 \& 97-BLG-56).

We study stars by the radiation emergent from their surface layers. While
the theory of numerical radiative transfer has made great progress over the
last decade, its application to stellar atmospheres has been, for the most
part, limited to one-dimensional equilibrium models. One reason was, partly,
because of technology $-$ we were unable to observe directly the surfaces of
other stars. 
Stellar interferometry, eclipsing binaries, and Doppler imaging, provide
valuable surface information, like limb darkening and stellar spots. 
However, by its nature, microlensing imaging allows us to observe the
center-to-limb variation in spectral lines (and their profiles). Such
observations are invaluable for the construction of detailed models of
the atmosphere, as it has been long known from the study of the Sun.
The solar disk brightness distribution as a
function of wavelength has been mapped into the depth distribution of
temperature and density, which allows a complete synthesis of the observed
solar spectrum (Kurucz 1990). 
The situation with stars has been confined to very much 
lower accuracy; the data cannot be used to build a model atmosphere, as in the
case of the Sun.

The 2-D and 3-D structure of stellar atmospheres other
than the Sun is not well known. Theoretical modelling relies heavily on
the solar atmosphere model (Nordlund \& Dravins 1990), 
but this situation is unfortunate for a number of reasons. There has been an
increased demand for accurate stellar models from a number of fields,
especially regarding red giants and supergiants. 
Our Sun is not a good zero point for most of these
applications.

\section{Gravitational Microlensing as a Stellar Microscope}

The amplification of a point source by 
a point mass, $M$, depends only on their
projected separation $d$~ (Paczy\'nski 1986),
\beq \label{amp}
A(d)= {d^2+2 \over d(d^2+4)^{1/2}}
\eeq
where $d$ is expressed in units of the angular radius of the
Einstein ring of the 
lens, $\theta_{_E}=([4GM/c^2][D_{_{\rm LS}}/D_{_{\rm OL}}
D_{_{\rm OS}}])^{1/2}$, and $D_{_{\rm OL}}, D_{_{\rm LS}}$,
and $D_{_{\rm OS}}$ are the distances between the observer, lens 
and source. The total flux received from an extended
source is therefore obtained by integration over its 
infinitesimal elements,
\beq \label{extended}
F(t)=\int_0^{\tR} r dr B(r) \int_0^{2\pi} d\theta A(d),
\eeq
where $B(r)$ is the surface brightness profile
of the source in the projected polar coordinates $(r,\theta)$
around its center. 
For simplicity, we assume that the stellar
emission is circularly symmetric, although this is not a limitation
(see Heyrovsky \& Loeb 1997; Heyrovsky, Loeb, \& Sasselov 1997),
and denote the source radius by $R_s=\tR\times (\theta_{_E}D_{_{\rm OS}})$.
All projected length scales
are normalized by the Einstein ring radius.
Using
the law of cosines
\beq \label{uexp}
d= \vert{\vec d_0}-{\vec r}\vert=
(d_0^2 -2 d_0r\cos\theta+r^2)^{1/2}~,
\eeq
where $d_0=(b^2+ v^2t^2)^{1/2}$ is the projected 
separation between the lens and the source center,
$b$ is the projected impact parameter of the source center,
$v$ is the transverse velocity of the lens
in units of $\theta_{_E}D_{_{\rm OL}}$, 
and $t=0$
is the midpoint time of the lensing event. For a circularly
symmetric source,
the resulting light curve is time-symmetric with $F(t)=F(-t)$. 
 
As a result of this dependence on $B(r)$, when the source is resolved,
$F(t)$ is wavelength dependent through $B(r)$, despite the achromatic nature
of gravitational lensing. For example, in the optical continua, $B(r)$ will
reflect the wavelength dependence of limb darkening, hence broad-band
photometry will show chromaticity due to that effect (Gould \& Welch 1996;
Heyrovsky, Loeb, \& Sasselov 1997; Valls-Gabaud 1997). 
Narrow-band photometry could reveal
resonant line emission in cool giants in a much more dramatic ways, as
shown by Loeb \& Sasselov (1995). And finally, high-resolution spectra
will reveal spectral line profile changes with time, representing the
subtle center-to-limb variation of different atomic and molecular
transitions.

\section{Probing the Atmospheres of Red Giants}
\subsection{Stellar Microscopy}

\begin{figure}[tb]
\plottwo{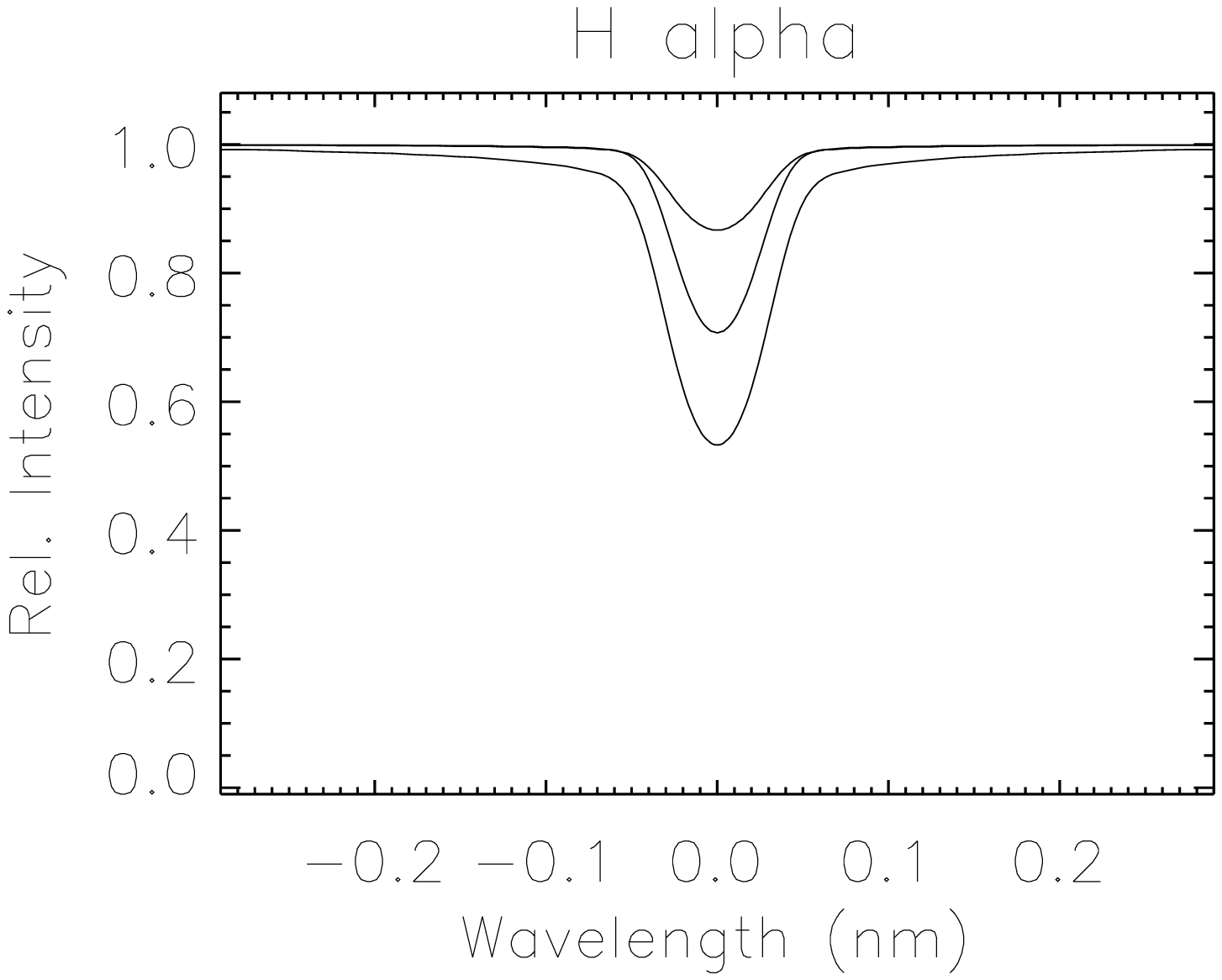}{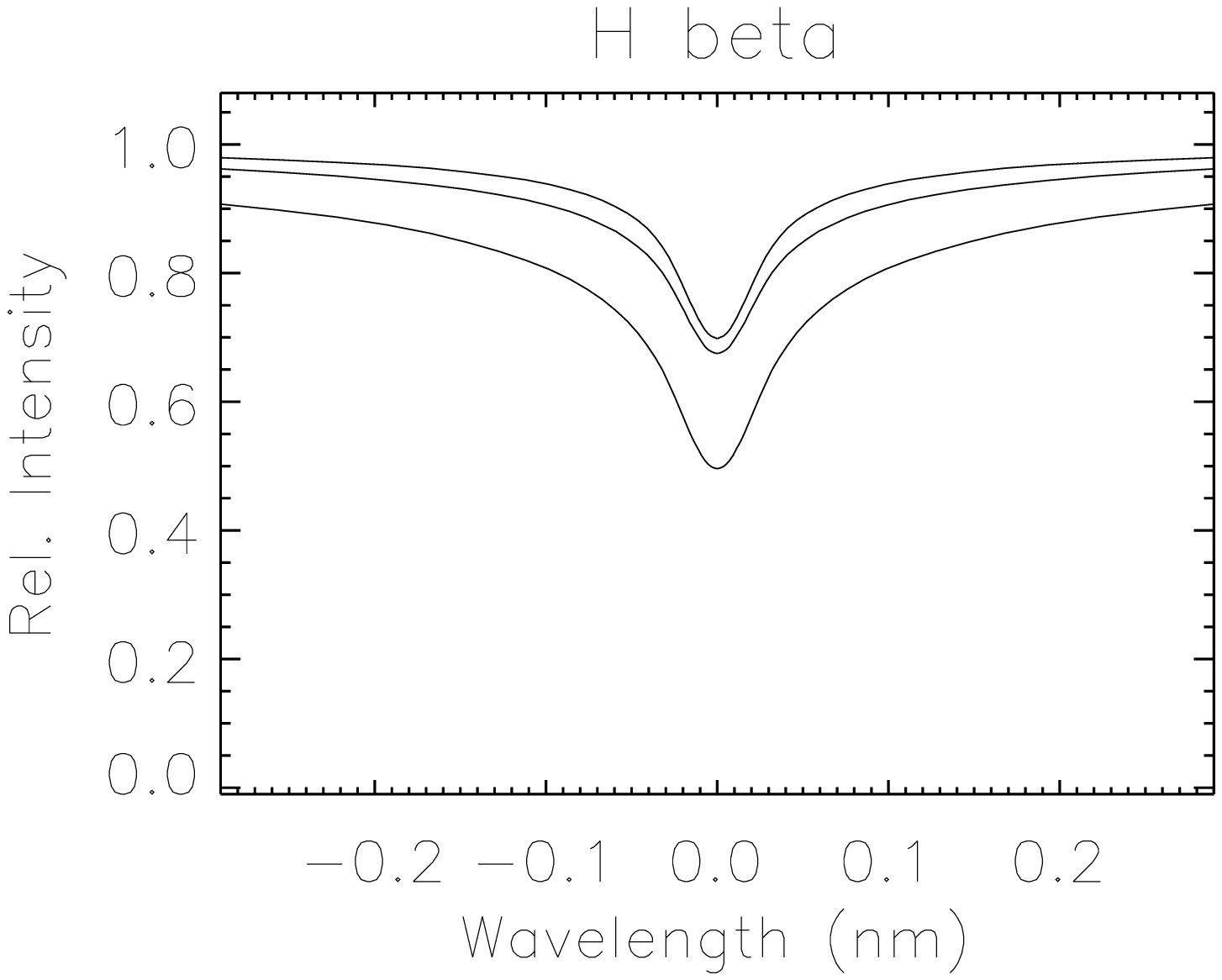}
\caption{Line profiles at three different viewing angles (characterized by
their cosines $\mu$) on the stellar disk
from the non-LTE model of the M95-30 red giant (Sasselov 1997a). 
a) Three profiles of 
the hydrogen H$\alpha$ line from disk center at
$\mu$=0.98 (bottom), through $\mu$=0.5, to very near the limb at $\mu$=0.02 
(top).
b) Same for the hydrogen H$\beta$ line. Note the difference in profile shape
between the two lines,
$as~well~as$ their change with $\mu$.}
\end{figure}
 
Stellar disks are projected hemispheres, which implies an axisymmetric
variation of the intensity with position as it maps into a variation with
depth. In general, the emergent continuum radiation at the center of the disk
is formed deeper than the radiation we see near the limb. Over a very wide
spectral range for stars of different temperature, the center-to-limb variation
of the continuum emission is manifested as limb darkening.
Unlike continuum radiation, spectral lines provide a larger choice of
center-to-limb variations. In the majority of stars we deal with absorption
lines of different strength, arising from atomic or molecular bound-bound
transitions. Most moderately strong and weak lines diminish towards the limb
(Figure 1). This variation could be observed in a microlensing event
by measuring the total equivalent widths of spectral lines on medium-resolution
spectra with high S/N (as shown in Figure 3), or directly $-$ on 
high-resolution spectra. The change of the H$\alpha$ line seen in Fig. 3,
reflects the expected change in the line at different viewing angles
$-$ strong at disk center (lens position 0.0) and weak at disk limb (lens
position 1.0).
 
The variation of intensity from disk center to disk limb is studied 
theoretically
by computing the contribution function of each transition; a function which
defines the line forming region in space (i.e. depth in atmosphere) and
frequency. Here we define (see Magain 1986) the non-LTE contribution
function, CF, for the relative line depression, as:
$$
{CF(lg~\tau _0)~=~{{ln~10}\over \mu}{\tau}_0{\kappa _l \over \kappa _0}
\Bigl(1-{{S_l}\over I_c}\Bigr) e^{-{{\tau _R}\over \mu}}},
$$
where ${\tau}_0$ is the optical depth at a reference wavelength ${\lambda}_0$,
${\tau}_R$ is defined in terms of ${\kappa}_R = {\kappa}_l +{\kappa}_c S_c/I_c$$
{\kappa}_0$ is the absorption coefficient at ${\lambda}_0$, ${\kappa}_l$
and ${\kappa}_c$ are the line and continuum absorptions, respectively,
S$_l$ and S$_c$ are the line and continuum source functions, respectively,
I$_c$ is the emergent continuous intensity (if the line were absent), and
$\mu$~=~cos~$\theta$. The traditional dependence on frequency has been omitted
from our notation for simplicity.  The CF for a given synthesized spectral
line will indicate where the line is formed, $i.e.$ the line forming region,
and consequently, at what velocity. We show the CFs for two hydrogen lines
in our M95-30 model in Figure 2.
 
\begin{figure}[tb]
\plottwo{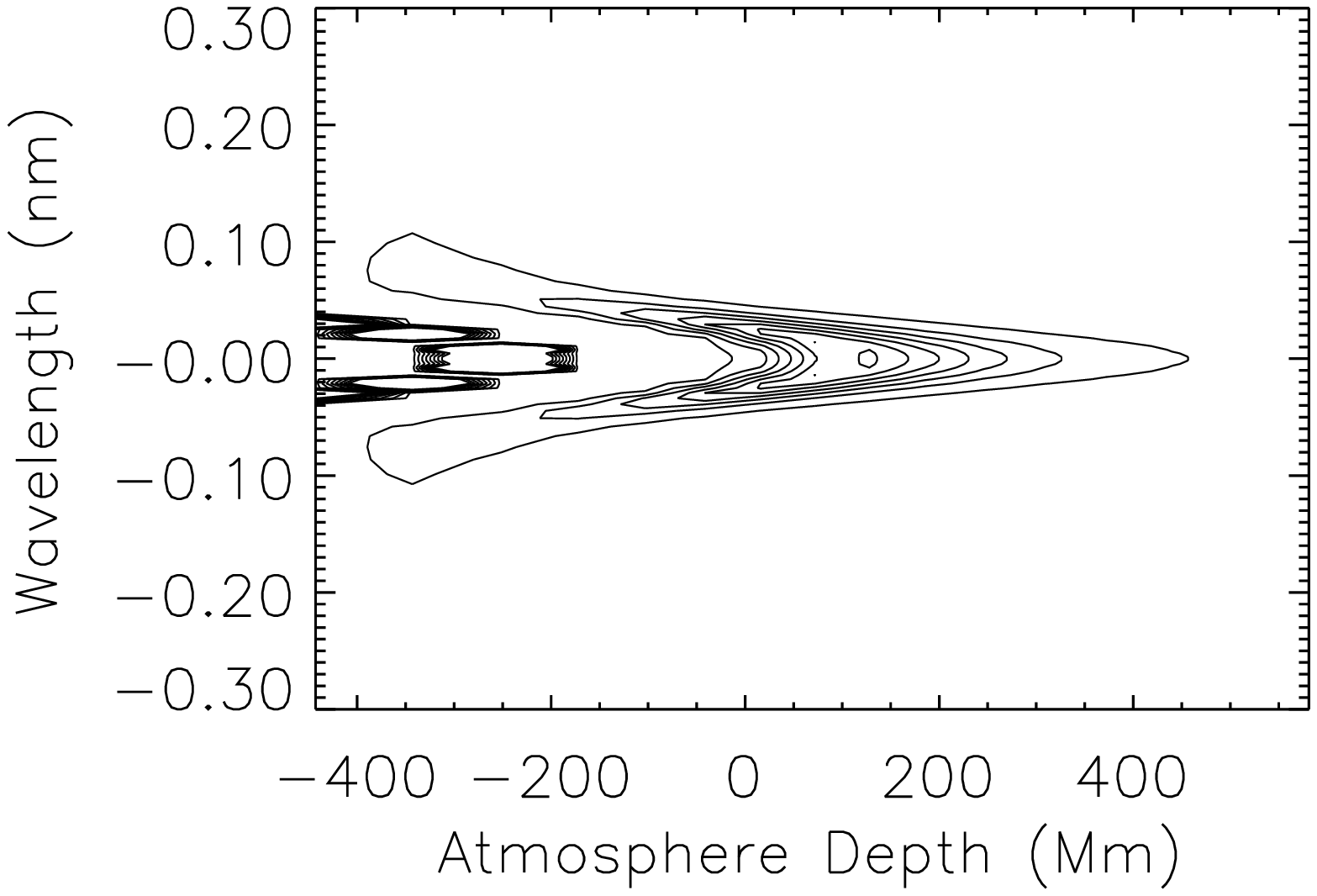}{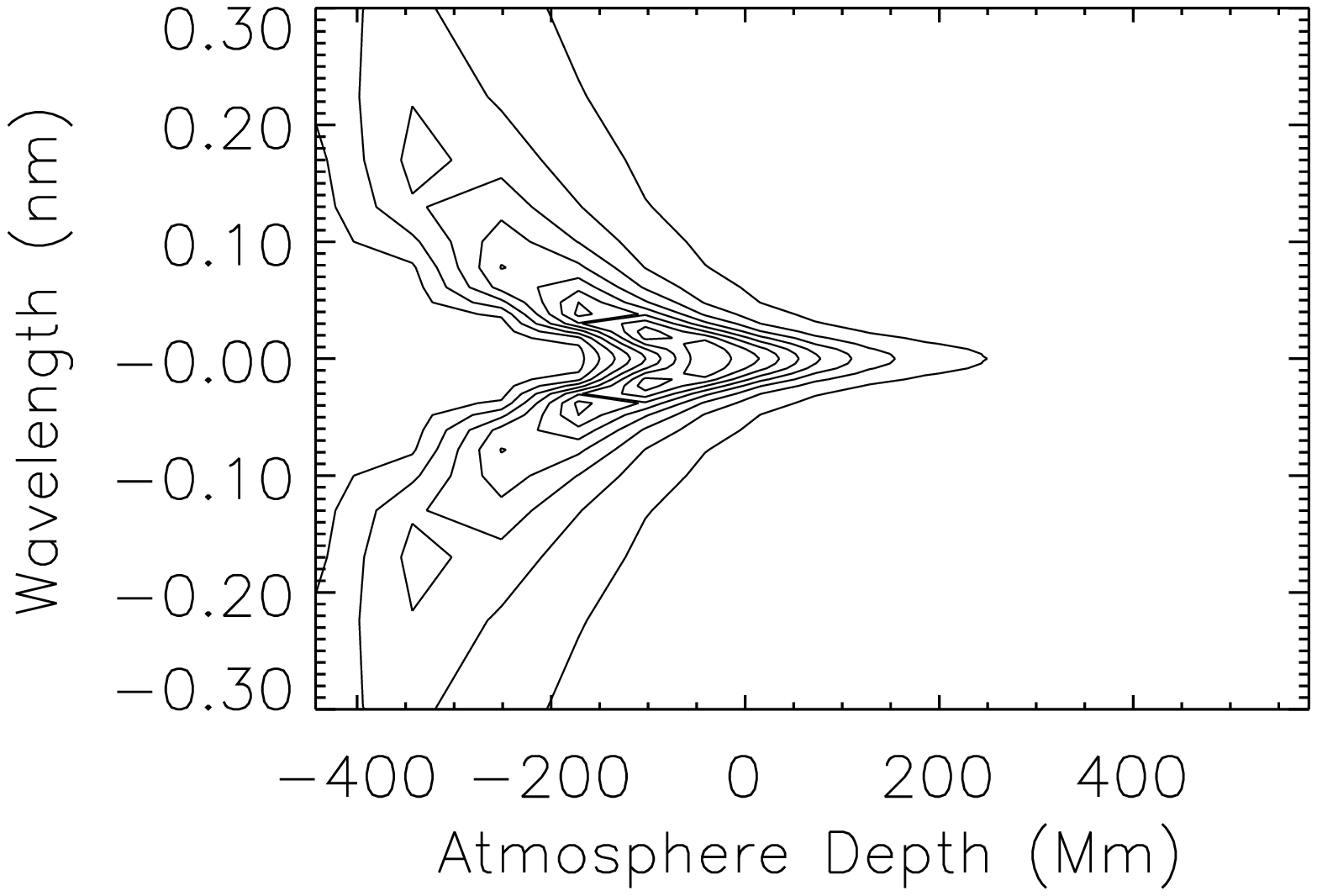}
\caption{Line forming regions, represented by the CFs, of the hydrogen H$\alpha$
and H$\beta$ lines shown in Figs.1ab. The contribution functions (CFs) are in
units of intensity and are angle averaged. a) The hydrogen H$\alpha$ line
is formed higher in the atmosphere overall, especially its line core. b) The
hydrogen H$\beta$ line is formed deeper and has well developed wings. However,
due to the background line and continuum opacity these line wings will not
be visible at small viewing angles. Both lines will have strong, but different,
center-to-limb variation.}
\end{figure}
 
Microlensing suffers from some of the same degeneracies inherent to Doppler
imaging, namely with respect to the lens path. It could be overcome in the
coming years with microlensing parallax observations (using spacecraft on
the Earth orbit). However, unlike Doppler imaging, line profile changes are
easy to measure and interpret. Of course, the rate of rotation is of no
importance to microlens imaging - in fact, microlensing can be used to
determine $v$sin$i$ independently (Maoz \& Gould 1994) from the Doppler
shifts of the spectral lines.
 
In order to explore the complete optical spectrum of a lensed star with
all necessary detail, one needs a very fast code which ``microlenses" 
all frequencies/wavelengths of an extended source as a function of time.
Such an efficient method was developed by Heyrovsky \& Loeb (1997);
Heyrovsky, Loeb, \& Sasselov (1997), and is used in the calculations 
of Fig.3.

By now there are more than 150 microlensing events 
detected towards the Galactic bulge by the MACHO and OGLE2 projects.
The standard light curve of an event
is characterized by two observables: its 
peak amplitude and duration (Paczy\'nski 1986).
Finite-size effects arise from resolved features on the surface of the 
lensed star.
Finite-size effects have been studied as methods to partially
remove the degeneracy of microlensing light curves through the alterations
of the standard light curve (Nemiroff \&
Wickramasinghe 1994; Gould 1994, 1995; Witt \& Mao 1994; Gould \& Welch 1996),
its polarized emission (Simmons, Newsam, \& Willis 1995),
spectral shifts due to stellar rotation (Maoz \& Gould 1994),
and narrow-band photometry in resonance lines (Loeb \& Sasselov 1995).
Here we want to put the emphasis on the inverse problem, that of
stellar microscopy $-$ reconstructing
the stellar surface features and probing the stellar atmosphere.
 
Red giants comprise a major component of the stars that are being monitored
in the bulge by the microlensing surveys. For example, in the OGLE dataset
about 19\% of all the stars have radii $\gsim 5R_{\odot}$ (Loeb \& Sasselov
1995). Therefore, for $D_{_{\rm OL}}=6~{\rm kpc}$,
the fraction of giant lensing events with $b< \tR$ (when the lens transits
across the disk of the source star)
is of order $\langle\tR\rangle \approx 4\%\times (M/0.1M_\odot)^{-1/2}$.
This is a large fraction, and observations are additionally aided by the
intrinsic brightness of the giants.
 
\begin{figure}[h]
\includegraphics{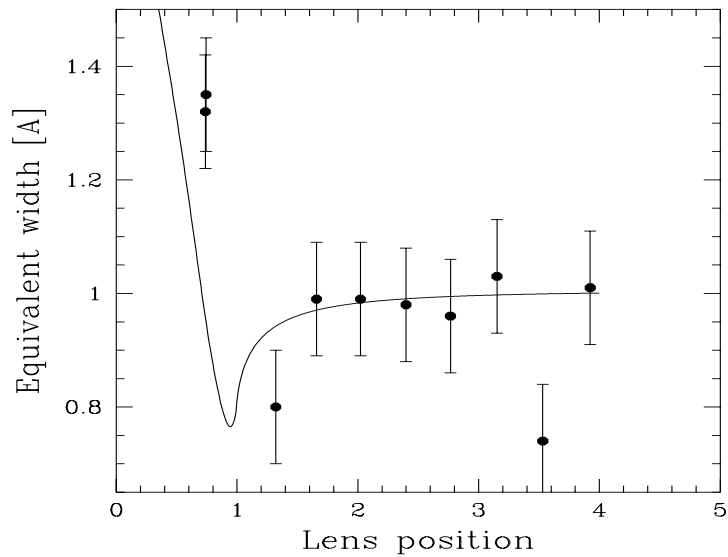}
\vspace*{3.4in}
\caption{The theoretical model variation of the H$\alpha$ line (solid
line) compared to the observed variation (equivalent width) in M95-30
by Alcock et al. (1997). Unfortunately, there were no observations made
near the limb crossing nights (lens position 1.0). The error bars may
be an underestimate, given the quality of the spectra and the confused
spectral reagion around the H$\alpha$ line. The observations between
lens positions 1.4 and 4.0 were obtained on 8 consecutive nights (from
Heyrovsky, Sasselov, \& Loeb 1998).}
\end{figure}
 
Stellar microscopy can constrain dramatically models of giant atmospheres
by providing center-to-limb variations of different line profiles in the
spectra of bulge red giants. Gravitational microlensing
offers an easily accessible, immediate, and inexpensive way to image at
least some types of stars. It also offers access to stellar
populations (in the Galactic bulge and Magellanic Clouds), which are beyond
the reach of any interferometer.
Gravitational lensing does resolve, $as~well~as$ amplify, thus making
17 $mag$ bulge giants as bright as 12 $mag$ $-$ a target easily accessible
to echelle spectrographs.

There is already a proof-of-concept result, based on the spectroscopy obtained
by the MACHO-GMAN collaboration (Alcock et al. 1997) of event M95-30.
Our 1-D model of the lensed red
giant in that event reproduces well the observations of the H$\alpha$ line
shown in Figure 3, as well as the TiO bands and the photometry (Heyrovsky,
Sasselov, \& Loeb 1998). We are currently working on the high-quality
spectra of M95-30 obtained by Alcock et al. (1997) to improve the model.

\subsection{Challenges in Modelling of Cool Giants}

Red giants are a major component of any
galactic stellar population. The parameters of the red giant branch
are derived from multi-color photometry interpreted with stellar atmosphere models
(e.g. Vandenberg \& Bell 1985; Kurucz 1992).
Such model atmospheres are one-dimensional 
integrations in semi-infinite slabs in hydrostatic and radiative equilibrium.
Many problems and the need to improve these models were reviewed recently
by Kurucz (1997). Many assumptions and simplifications are being made
$-$ some of them have not been verified by direct observations. Such examples are
the treatment of convective transport and small-scale velocity fields (e.g.
micro- and macro-turbulence).

In traditional one-dimensional radiative transfer modeling one usually
parameterizes small-scale dynamic motions with an assumed microturbulent
velocity distribution.
Small-scale in this case refers to motions that occur within a photon
mean free pathlength, so that these motions contribute to the
intrinsic broadening of the line profile.
On top of these small-scale motions there are motions on larger scales
that have to be accounted for in observations with limited spatial
resolution.
These are often called macroturbulence, which is a mis-characterization,
since most of these motions occur in well organized forms like radial-
and non-radial pulsations, and convective flows.
Macroturbulence is applied as additional broadening to the emergent
line profiles.
As such it has no consequences for the radiative transport and
energy balance in the atmosphere.
Microturbulent broadening, on the other hand, does 
affect the flow of radiation through the stellar atmosphere
because they provide additional line opacity.

Although one-dimensional models exist for stars with a wide variety
of effective temperatures and gravities (e.g., the Kurucz 1992 models
or the Allard \& Hauschildt 1995 models for later type stars) little
modeling has been done of the multi-dimensional structure of stars
other than the Sun (Dravins 1990), and practically none for red giants.
The stellar evolution models on the other hand are dependent on a number
of parameters, which describe poorly understood physics, e.g. the
treatment of convection near the surface of the star. Using scaling
arguments, Schwarzschild (1975) suggested that red giants may have only
a small number of very large convection cells on their surface.
High-resolution microlensing data can provide insights into these
issues and help develop efficient and physically consistent recipes for calculating
small-scale velocity fields.
For example, assuming that small-scale velocity fields
are well represented by a turbulent spectrum one could model the 
depth-dependent microturbulent velocity with a simple self-consistent model
of turbulence in the spirit of Canuto, Goldman, \& Mazzitelli (1996),
and treat the subgrid cascade. It is clear, however, that a 3-D transfer
calculation is necessary to provide the initial spectral scales for such a
scheme (Sasselov 1997b). 

Problems such as described above have hindered significant improvement
in the determination of fundamental stellar properties over the past
20 years.
Just to illustrate one of the current problems involving red giants 
calibrations, consider the ``age of the Universe" debate.
Calculating isochrones for globular clusters $-$ using the same evolutionary
models and assumptions, but different model atmospheres for $T_{\rm eff}$
and bolometric correction calibrations (Bell-Gustafsson vs. Kurucz), leads
to 20\% difference in age estimates (Salaris et al. 1997; Vandenberg 1997).
Given the conflicting results in interpreting the HIPPARCOS satellite data
on subdwarfs calibrations of globular cluster sequences (Reid 1997; Pont
1997), the issue of isochrone calibration remains open and very important.
The need for 2-D and 3-D tools for atmosphere modelling comes also from recent
successful direct stellar imaging with the Hubble Space Telescope
(e.g. Uitenbroek, Dupree, \& Gilliland 1997).

\section{Conclusions}

Gravitational microlensing offers unique ways to:  (1) study depth dependence
in red giants atmospheres of line formation, small-scale velocity fields,
and convective motions; (2) estimate occurence of surface features like
spots and active regions in red giants; and (3) derive rotation ($v$sin$i$)
and limb darkening in the bulge population of red giants. The imaging of
spots and active regions suffers from degeneracies similar to those in
Doppler imaging. However the ability of microlens imaging
to resolve center-to-limb variations in spectral lines is exceptional and
unique. So is the value of such information to the construction of a new
generation of giants model atmospheres.
\bigskip

It is a pleasure to acknowledge my collaborators - D. Heyrovsky and A. Loeb, and
to thank I. Shapiro for his support of this work.


\begin{references}

\reference Alcock, C., et al. 1997, ApJ, in press (astro-ph/9702199)
\reference Allard, F. \& Hauschildt, P., 1995, ApJ, 445, 433
\reference Armstrong, J.T. et al. 1995, {\it Physics Today}, 48, 42
\reference Canuto, V. M., Goldman, I., \& Mazzitelli, I. 1996, ApJ, 473, 550
\reference Dravins, D. 1990, A\&A, 228, 218
\reference Heyrovsky, D. \& Loeb, A. 1997, ApJ, in press (astro-ph/9702097)
\reference Heyrovsky, D., Loeb, A., \& Sasselov, D. 1997, in
{\it Variable Stars and the Astrophysical Returns of Microlensing
Surveys}, R. Ferlet, J.P.Maillard, \& B. Raban, eds., Editions Frontieres, p.417
\reference Heyrovsky, D., Sasselov, D., \& Loeb, A. 1998, in preparation
\reference Gould, A. 1992, ApJ, 392, 442
\reference -----------.1994, ApJ, 421, L71
\reference -----------.1995, ApJ, 441, L21
\reference Gould, A., \& Welch, D. 1996, ApJ, 464, 212
\reference Kurucz, R.L. 1990, in {\it Atomic Spectra for Astroph. \& Fusion},
Hansen,J.E., ed., North-Holland, p.20
\reference Kurucz, R.L. 1992, in {\it Stellar Population of Galaxies},
Barbuy,B.,\& Renzini,A., eds., Kluwer, p.225
\reference Kurucz, R.L. 1997, in 
{\em Fundamental Stellar Properties: The Interaction between
Observation and Theory}, T.R.Bedding et al., eds., Kluwer, (CfA preprint 4571)
\reference Lennon, D.J., Mao, S., Fuhrmann, K., \& Gehren, T. 1996, ApJ, 471, L23
\reference Loeb, A., \& Sasselov, D.D. 1995, ApJ, 449, L33
\reference Magain, P. 1986, A\&A, 163, 135
\reference Maoz, D., \& Gould, A. 1994, ApJ, 425, L67
\reference Nemiroff, R.J., \& Wickramasinghe, W.A.D.T. 1994, ApJ, 424, L21
\reference Nordlund, A. \& Dravins, D. 1990, A\&A, 228, 155
\reference Paczy\'nski, B. 1986, ApJ, 304, 1
\reference Pont, F. 1997, in {\em IAU General Assembly - JD14}, Kyoto
\reference Reid, I.N. 1997, in {\em IAU General Assembly - JD14}, Kyoto
\reference Salaris, M., Degl'Innocenti, S., \& Weiss, A. 1997, ApJ, 484, 986
\reference Sasselov, D.D. 1996, in {\it Cool Stars 9}, 
Pallavicini,R. \& Dupree,A.K., eds., ASP Conf Ser. 109, p.541
\reference Sasselov, D.D. 1997a, in
{\it Variable Stars and the Astrophysical Returns of Microlensing
Surveys}, R. Ferlet, J.P.Maillard, \& B. Raban, eds., Editions Frontieres, p.141
\reference Sasselov, D.D. 1997b, in
{\em Fundamental Stellar Properties: The Interaction between
Observation and Theory}, T.R.Bedding et al., eds., Kluwer, p.253 (astro-ph/9705023)
\reference Schwarzschild, M. 1975, ApJ, 195, 137
\reference Simmons, J.F.L., Newsam, A.M., \& Willis, J.P. 1995, MNRAS, 276, 182
\reference Uitenbroek, H., Dupree, A.\ K., Gilliland, R.\ L.\ 1997,
in J.\ Bookbinder and R.\ Donahue (eds.), Cool Stars, Stellar Systems,
and the Sun, 10th Cambridge Workshop, this volume
\reference Valls-Gabaud, D. 1997, MNRAS, in press (astro-ph/9708098)
\reference Vandenberg, D., \& Bell, R.A. 1985, ApJS, 58, 561
\reference Vandenberg, D. 1997, in {\em Fundamental Stellar Properties:
The Interaction between Observation and Theory}, T.R.Bedding et al., eds., Kluwer.
\reference Witt, H. J., \& Mao, S. 1994, ApJ, 430, 505

\end{references}
\end{document}